\newcommand{\nc}{\newcommand}
\nc{\beq}{\begin{equation}}   \nc{\eeq}{\end{equation}}
\nc{\bea}{\begin{eqnarray}}   \nc{\eea}{\end{eqnarray}}
\nc{\baa}{\begin{array}}      \nc{\eaa}{\end{array}}
\nc{\bit}{\begin{itemize}}    \nc{\eit}{\end{itemize}}
\nc{\ben}{\begin{enumerate}}  \nc{\een}{\end{enumerate}}
\nc{\bce}{\begin{center}}     \nc{\ece}{\end{center}}
\def\beqa{\begin{eqnarray}}
\def\eeqa{\end{eqnarray}}

\def\MPL #1 #2 #3 {{\sl Mod.~Phys.~Lett.}~{\bf#1} (#3) #2}
\def\NPB #1 #2 #3 {{\sl Nucl.~Phys.}~{\bf #1} (#3) #2}
\def\PLB #1 #2 #3 {{\sl Phys.~Lett.}~{\bf #1} (#3) #2}
\def\PR #1 #2 #3 {{\sl Phys.~Rep.}~{\bf#1} (#3) #2}
\def\PRD #1 #2 #3 {{\sl Phys.~Rev.}~{\bf #1} (#3) #2}
\def\PRL #1 #2 #3 {{\sl Phys.~Rev.~Lett.}~{\bf#1} (#3) #2}
\def\RMP #1 #2 #3 {{\sl Rev.~Mod.~Phys.}~{\bf#1} (#3) #2}
\def\ZPC #1 #2 #3 {{\sl Z.~Phys.}~{\bf #1} (#3) #2}
\def\IJMP #1 #2 #3 {{\sl Int.~J.~Mod.~Phys.}~{\bf#1} (#3) #2}
\def\NIM #1 #2 #3 {{\sl Nucl.~Inst.~and~Meth.}~{\bf#1} {#3} #2}
\def\alp{\alpha^\prime}
\def\etc{{\it etc.}}
\def\br{BF}

\def\ocp{{\cal O}_{CP}}

\def\h{h}
\def\mh{m_{\h}}

\def\eg{{\it e.g.}}

\def\epem{e^+e^-}

\def\tauptaum{\tau^+\tau^-}

\def\lsim{\mathrel{\raise.3ex\hbox{$<$\kern-.75em\lower1ex\hbox{$\sim$}}}}
\def\gsim{\mathrel{\raise.3ex\hbox{$>$\kern-.75em\lower1ex\hbox{$\sim$}}}}
\def\@versim#1#2{\vcenter{\offinterlineskip
        \ialign{$\m@th#1\hfil##\hfil$\crcr#2\crcr\sim\crcr } }}

\def\ie{{\it i.e.}}

\def\gam{\gamma}

\def\anti{\overline}

\def\fbi{~{\rm fb}^{-1}}

\def\gev{\,{\rm GeV}}

\def\hl{h^0}
\def\hh{H^0}
\def\ha{A^0}

\def\mhh{m_{\hh}}
\def\mha{m_{\ha}}

\def\tanb{\tan\beta}

\def\mt{m_t}

\def\mw{m_W}

\def\h{h}
\def\mh{m_{\h}}

\documentstyle[12pt,equations,epsf]{article}
\def\ie{{\it i.e.}}

\def\9{\phantom 0}      %%% for lining up numbers in columns
\renewcommand\linebreak{\unskip\break} %% breaks line & still justifies
\begin{document}
\newlength{\captsize} \let\captsize=\small % use \let\normalsize=\captsize
\newlength{\captwidth}                     % just before \caption{ ...
%%%%%%%%%%%%%%%%%%%%%%%%%%%%%%%%%%%%%%%%%%%%%%%%%%%%%%%%%%%%%%

%\preprint{
%
\font\fortssbx=cmssbx10 scaled \magstep2
\hbox to \hsize{
%
%\special{psfile=uwlogo.ps
% hscale=8000 vscale=8000
% hoffset=-12 voffset=-2}
%\hskip.5in \raise.1in
%
$\vcenter{
\hbox{\fortssbx University of California - Davis}
%\hbox{\fortssbx University of Wisconsin - Madison}
}$
\hfill
$\vcenter{
\hbox{\bf UCD-98-10} 
\hbox{\bf IFT-UW-98-26}
\hbox{\bf hep-ph/9809306}
%\hbox{\bf MADPH-95-884} 
%\hbox{\bf IUHET-299}
\hbox{September, 1998}
}$
}
%}

%
\medskip
\begin{center}
\bf
DETERMINING THE RELATIVE SIZE OF THE CP-EVEN AND CP-ODD
HIGGS BOSON COUPLINGS TO A FERMION AT THE LHC
\\
\rm
\vskip1pc
{\bf John F. Gunion$^a$ and Jacek Pliszka$^{a,b}$}\\
\medskip
{\em a) Davis Institute for High Energy Physics, 
University of California, Davis, CA, USA}\\
{\em b) Institute for Theoretical Physics, Warsaw University, Warsaw, Poland}\\
\end{center}

\begin{abstract}
We demonstrate that the relative size of the CP-even and CP-odd couplings
of a Higgs boson to $t\anti t$ 
can be determined at the LHC at a level that is very useful in
discriminating between models, provided that a Higgs signal
can be observed at a reasonable $S/\sqrt B$ level
in the $t\anti t\h$ (with $\h\to\gam\gam$ or $b\anti b)$ channels. 
In particular,
the CP-even nature of a SM-like Higgs boson's coupling to $t\anti t$
can be confirmed with very substantial statistical significance
using the $t\anti t \h$ ($\h\to \gam\gam$) production/decay mode.
The key to achieving good discrimination between models is to make full use
of the difference in the final state distributions for
different ratios of the CP-even to CP-odd couplings. For our analysis,
we employ the very convenient and simple optimal analysis procedures
derived previously. 

\end{abstract}

\section{Introduction}

It will be very important to directly determine the CP nature of any Higgs
boson ($\h$) that is discovered. Since it is almost certain
that a substantial number of Higgs boson events will first be
available at the LHC, it will be very desirable to use LHC
data to verify the CP nature of any observed Higgs without waiting for an
$\epem$ collider. The only procedure 
with a reasonable level of viability proposed to date is to
employ weighted integrals of the $pp\to t\anti t\h X$ final state~\cite{gunhe}.
These were shown to provide significant 
ability to establish that a SM-like Higgs boson has a purely CP-even
$t\anti t$ coupling.
Subsequently, procedures for optimizing the extraction 
of the relative magnitude of different components of a cross section
were developed and their power demonstrated
when applied to determining the
relative magnitude of the CP-even and CP-odd $t\anti t\h$ couplings 
in the process $\epem\to t\anti t\h$~\cite{optimal}.
Here, we apply these same techniques to the $t\anti t \h X$ final state
at the LHC and demonstrate substantial improvement as compared to the
weighted integral technique of Ref.~\cite{gunhe}. In general,
a very meaningful determination of the relative magnitude of the
CP-even and CP-odd $t\anti t$ couplings of the $\h$ will be possible
provided only that a signal with reasonable statistical significance
is present. The key to the optimal analysis techniques is to take
full advantage of all the information
available in the cross section as a function of the kinematical variables.

For our analysis, it is important that a relatively 
pure sample of $t\anti t F X$ events can be isolated
for one or more final states $F$ into which the Higgs decays. Here, we will 
consider the $F=\gam\gam$ and $F=b\anti b$
final states originally discussed in Refs.~\cite{ttgamgam} and \cite{dgv},
respectively. The events will
consist of $t\anti t\h X$ ($\h\to F$) signal events and background events.
(For a light $\h$, our focus here, interference between signal and background
amplitudes can be neglected.)
The $pp\to t\anti t F X$ cross section takes the form
\beq
\Sigma(\phi)\equiv {d\sigma\over d\phi}
=C\left[a^2 f_a(\phi)+b^2f_b(\phi)\right]+f_B(\phi)\,,
\label{sigform}
\eeq
where $\phi$ denotes the final state phase space configuration,
$C$ is a coefficient determined by experimental efficiencies, $\br(\h\to F)$
\etc, and $a$ and $b$ are the CP-even and CP-odd Higgs couplings
appearing in the Feynman rule,
\begin{equation}
t\anti t\h:~~-\anti t(a+ib\gamma_5) t{g\mt\over 2\mw}\,.
\label{coupdefs}
\end{equation}
(The SM Higgs boson 
has $a=1$ and $b=0$.  A purely CP-odd Higgs boson has $a=0$ and $b\neq 0$.)
We note that there is no $ab$ cross term in Eq.~(\ref{sigform})
since we do not consider any observable sensitive
to the spin directions of the top quarks. Further, since both the $a^2$
and $b^2$ terms are of a CP-conserving nature, the functions $f_a$ and
$f_b$ do not change when the $t$ and $\anti t$ momenta are interchanged.

In Eq.~(\ref{sigform}), $f_{a,b}(\phi)$ are, in principle, precisely known
functions of $\phi$ and have well-determined relative normalization.
The only possible uncertainties in the $f_{a,b}$ arise through their
dependence on quark/gluon distribution functions; we will assume
that the relevant distribution functions will have been determined
with sufficient precision, using other LHC data and data from 
other accelerators, that these uncertainties can be neglected.
We also assume that the background contribution $f_B(\phi)$ will
be precisely determined using experimental data outside but near 
the Higgs mass peak in the final state $F$. This precise measurement
will include an accurate determination of both the functional
form and the normalization, including
all experimental efficiencies. Statistical fluctuations in the background
within the accepted Higgs mass bin in the state $F$ are automatically
incorporated in the procedure we will employ.

\section{The Optimal Analysis Procedure and Results}

As in Ref.~\cite{gunhe}, we wish to discriminate between models purely
on the basis of the difference between the phase space distribution
dependence of $f_a$ and $f_b$,
not relying on knowledge of $C$. To this end, it will be convenient to
define the rescaled functions $\hat f_{a,b,B}(\phi)\equiv 
f_{a,b,B}(\phi)/I_{a,b,B}$
where $I_{a,b,B}\equiv\int f_{a,b,B}(\phi)d\phi$ 
so that $\int \hat f_{a,b,B}=1$. 
(The $\phi$ integral will be restricted by appropriate cuts.)
% $r=0.288$.
Defining $r\equiv I_b/I_a$ and $\alpha\equiv a^2/(a^2+rb^2)$, we 
may rewrite Eq.~(\ref{sigform}) as
\beq
\Sigma(\phi)= \sigma_S\left[\alpha
f_{\alpha}(\phi)+\hat f_b(\phi)\right]+\sigma_B\hat f_B(\phi)\,,
\label{sigp}
\eeq
where $f_{\alpha}(\phi)\equiv [\hat f_a(\phi)-\hat f_b(\phi)]$ ($\int
f_{\alpha}(\phi)d\phi=0$) and $\sigma_S=C(a^2I_a+b^2I_b)$ is the 
effective integrated signal cross section.
The background cross section is $\sigma_B=I_B$. Clearly, the ratio $\alpha$
is independent of the overall signal
normalization (although errors in $\alpha$ will certainly depend
on the signal and background rates). We now use the optimal analysis
technique of Ref.~\cite{optimal} to provide a determination
of $\alpha$ with the smallest possible error (in the Gaussian 
statistics approximation).

In general, for $\Sigma(\phi)=\sum_i c_if_i(\phi)+g(\phi)$
(where $g(\phi)$ and the $f_i(\phi)$ are known functions, 
including normalization)
the determination of the unknown signal coefficients $c_i$ 
with smallest possible statistical error is given by \cite{optimal}
\begin{equation}
c_i=\sum_k M_{ik}^{-1}I_k\,,\quad{\rm where}~~
M_{ik}\equiv \int {f_i(\phi)f_k(\phi)\over \Sigma(\phi)} d\phi\,,~{\rm and}~
I_k\equiv \int f_k(\phi) d\phi\,.
\label{ciform} 
\end{equation}
The covariance matrix for the $c_i$ is
\begin{equation}
V_{ij}\equiv \langle \Delta c_i\Delta c_j\rangle= 
{ M_{ij}^{-1} \sigma_T\over N}=M_{ij}^{-1}/L_{\rm eff}\,,
\label{cerror}
\end{equation}
where $\sigma_T=\int {d\sigma\over d\phi} d\phi$ is the integrated
cross section and $N=L_{\rm eff}\sigma_T$ is the total number of events,
with $L_{\rm eff}$ being the luminosity times efficiency.
This result is the optimal one regardless of the
relative magnitudes of the different contributions
to $\Sigma(\phi)$. In the Gaussian statistics limit,
it is equivalent to determining the $c_i$ by maximizing the
likelihood of the fit to the full $\phi$ distribution of all the events.
The increase in errors due to statistical fluctuations in
the presence of background are implicit in the 
background contribution to $\Sigma(\phi)$ appearing in Eq.~(\ref{ciform}),
which implies larger $M_{ij}^{-1}$ entries in Eq.~(\ref{cerror}).
From the result of Eq.~(\ref{cerror}), 
the $\chi^2$ in the $c_i$ parameter space is then computed as 
\beq
\chi^2=\sum_{i,j}(c_i^\prime-c_i)V_{ij}^{-1}(c^\prime_j-c_j)=
\sum_{i,j}(c^\prime_i-c_i)L_{\rm eff}M_{ij}(c^\prime_j-c_j)\,,
\label{chisqform}
\eeq
where, for our theoretical
analyses, the $c_i$ are the input model values for which $V_{ij}$ is
computed.\footnote{Once experimental data is available, one would
compute $M_{ij}$ using appropriate binning and actual data for $\Sigma(\phi)$.}

The above procedures are not altered if cuts are imposed
on the kinematical phase space over which one integrates; one simply restricts
all $\phi$ integrals to the accepted region.
If a subset, $\bar\phi$, of the kinematical
variables $\phi$ cannot be determined, then the optimal technique 
can be applied using the variables, $\hat\phi$, that {\it can} be observed
and the functions $\bar f_i(\hat\phi)\equiv \int f_i(\phi) d\bar\phi$.

We apply the above general technique to the specific case in hand
by identifying the signal function $f_1=\sigma_S f_{\alpha}$ and
the known function $g(\phi)=\sigma_S\hat f_b(\phi)+\sigma_B\hat f_B(\phi)$.
(The normalizations are fixed by the experimental measurements of
the signal and background rates, which include all efficiencies \etc)
The $\chi^2$ associated with a model yielding a value $\alpha^\prime$
that differs from the value $\alpha$ of the input model 
is then given by the following simple result:
\beq
\chi^2=(\alp-\alpha)^2\,{S^2\over B}\int d\phi {f_{\alpha}^2(\phi)\over
{S\over B}[\alpha f_\alpha(\phi)+\hat f_b(\phi)]+\hat f_B(\phi)}\equiv
(\alp-\alpha)^2{S^2\over B} [H(\alpha,S/B)]^2\,,
\label{chisqalpha}
\eeq
where the expected signal rate $S$ 
is that estimated (and eventually measured) for the input (true) model.
It is important to note that $H(\alpha,S/B)$ depends implicitly
on the channel $F$ being considered.
Defining the discrimination power, $D\equiv \sqrt{\chi^2}$, we have 
\beq 
D=|\alp-\alpha|{S\over \sqrt B} H(\alpha,S/B)
\label{ddef}
\eeq
Note that in the limit of small $S/B$,
\beq
H(\alpha,S/B)\to H\equiv \left[\int d\phi
\left\{f_\alpha^2(\phi)\over \hat f_B(\phi)\right\}\right]^{1/2}\,.
\label{hdef}
\eeq 

In applying the above to the $pp\to t\anti t F X$ process, 
we only consider the kinematical
variables associated with the $t\anti t F$ portion of the final state,
including its $\sqrt{\hat s}$ (where $\hat s$ is
the invariant mass-squared for the $t\anti t F$ combination)
and overall rapidity.
Any variables associated with the internal phase space of $X$ are 
inclusively summed over as usual (and fall into the $\bar\phi$ category
noted above). Also, distributions for the decay products of the $t$-quark
and the Higgs boson are not considered.

In order to establish that
we are, in fact, dealing with the $t\anti t F$ final state and completely
determine the kinematical point $\phi$ in phase space,
we must be able to reconstruct and identify the $t$ and $\anti t$.
However, identification of $t$
vs. $\anti t$ is not actually necessary since, as noted earlier,
our functions $f_a$ and $f_b$ do not change if the $t$ and $\anti t$
momenta are interchanged. Thus, in principle we can use
either the final state mode in which both the $t$ and $\anti t$
decay completely hadronically or that in which one has a leptonic $W$
decay and the other a hadronic $W$ decay.\footnote{The final 
state mode in which the $W$'s from both the $t$ and the $\anti t$
decay leptonically does not allow full determination of
the kinematics of the final state.} In this paper, we choose
to focus on the latter (mixed hadronic-leptonic) mode
as being possibly somewhat cleaner. However,
we anticipate that the purely hadronic final state could be used
to further increase our statistics.

In the final state mode in which one $t$ decays
leptonically and the other hadronically, there is a potential
two-fold ambiguity when determining the longitudinal momentum of the missing
neutrino by requiring that the leptonically-decaying $W$ be on-shell.
However, the wrong solution gives a reconstructed
top-quark mass that differs from the known value, $\mt$.
If we assume that energy resolutions and such will allow reconstruction
of $\mt$ to within $\pm 10\gev$ for the correct solution, then
we can identify the correct solution in those events for which
the incorrect solution gives a reconstructed top-quark
mass that differs by more than $10\gev$ from $\mt$. About 70\%
of the events that obey the global cuts specified below 
satisfy this criterion. We accept only such events.
The global cuts that we impose on the $t\anti t\h$
final state are such that there should be reasonable efficiency
for the required reconstructions. We require $|y_{t,\anti t,\h}|<4$.
A cut of $p_T>\mh/4$ is applied to the $\gam$'s or $b$'s in the $F$
final state in order to reduce the size of the background.
These are the same cuts as employed in Ref.~\cite{gunhe}.

We first give results for a Higgs mass of $\mh=100\gev$,
so as to allow direct comparison with Ref.~\cite{gunhe}. 
(We will discuss
later the sensitivity of our results to the $|y|$ cut and to $\mh$.)
For the above cuts and this mass, we find $r=0.288$. 
(Note that $r$ is independent of the final state channel $F$.) 

As already noted, we will consider the $F=\gam\gam$ and $F=b\anti b$
decay modes for the $\h$. The $F=\gam\gam$ mode will only
be considered in the case where the input model corresponds to
a SM-like $\h$, since it is only in this case that $\br(\h\to\gam\gam)$
is likely to be substantial. Following Ref.~\cite{ttgamgam}, 
the LHC ATLAS \cite{ATLAS} and CMS \cite{CMS} collaborations performed
detailed simulations of this channel 
as part of their technical design reports. Based on their results,
and assuming an
eventual integrated luminosity of $L=600\fbi$ (summing
over the two detectors), Ref.~\cite{gunhe} concluded
that one could expect $S\sim 130$ and $B\sim 21$ in the $F=\gam\gam$
mode for a SM-like $\h$ with mass of order $100$ GeV. 
For $F=b\anti b$, the best global cuts
and associated levels of signal and background have not
yet been firmly established.  We will adopt the procedure of Ref.~\cite{gunhe}
of examining results for several different extremes of $S/B$, assuming
a certain level of statistical significance $S/\sqrt B$. In this way,
we avoid making any specific assumption about the branching ratio for
$\h\to b\anti b$ decay, which is, in any case, absorbed into the overall
normalization factor $C$ of Eq.~(\ref{sigform}).
For both $F=\gam\gam$ and $F=b\anti b$, we will follow Ref.~\cite{gunhe}
and model the background cross section shape $f_B(\phi)$
[see Eq.~(\ref{sigform})] using the irreducible $pp\to t\anti t\gam\gam X$ 
and $t\anti t b\anti b X$ background processes, respectively.
In the case of $F=\gam\gam$, 
the ATLAS and CMS studies confirm the theoretical claims \cite{ttgamgam}
that the irreducible background should, indeed, be dominant. 
In the case of $F=b\anti b$, minimum-bias and other backgrounds may
enter \cite{rwf}. To the extent that the functional form
(in the kinematical variables $\phi$) of the sum of such backgrounds is
substantially different from the $f_B(\phi)$ calculated for the
irreducible backgrounds, the sensitivities that we compute in this paper
as a function of $S/B$ and $S/\sqrt B$
will have to be re-evaluated by the experimental collaborations.
We re-emphasize that in the actual experiment $f_B(\phi)$ will
be directly measured using data for which the $b\anti b$ mass is outside
the Higgs mass peak region.

In order to assess our ability to determine that a CP-even SM-like
Higgs boson is indeed CP-even, we will follow the procedure of
Ref.~\cite{gunhe} and consider three distinct Higgs coupling cases. We
compare [recalling that $\alpha=a^2/(a^2+rb^2)$]
\begin{itemize}
\item I) A Standard-Model-like Higgs boson, with $a\neq 0$ and $b=0$, \ie\ 
$\alpha_I=1$, to:
\item II) a CP-mixed Higgs boson, with $a=b$, \ie\ $\alpha_{II}=1/(1+r)$; and,
\item III) a  pure CP-odd Higgs boson, with $a=0$, $b\neq0$, \ie\ 
$\alpha_{III}=0$,
\end{itemize}
In order to compare with the results of Ref.~\cite{gunhe}, we
will compute the discrimination powers $D_{1,2}
=|\alpha_{II,III}-\alpha_I|(S/\sqrt B)H(\alpha_I,S_I/B)$ defined
in Eq.~(\ref{ddef}).
These are a measure of our ability to ascertain that a CP-even
SM-like Higgs boson is {\it not} CP-mixed or CP-odd, respectively.
In comparing, we use only the best-case weighting functions $\ocp$ considered in
Ref.~\cite{gunhe}. For $F=\gam\gam$ the best choice
for $\ocp$ was denoted by $\ocp=b_1$; for $F=b\anti b$ the best choice
was denoted by $\ocp=b_4$.

\begin{table}[h]
\caption[fake]{$t\anti t\gam\gam$ channel, $\mh=100\gev$: 
Discrimination powers $D_1$ and $D_2$ for distinguishing a SM-like Higgs
from a CP-mixed Higgs and from a CP-odd Higgs, respectively.
Results from Ref.~\cite{gunhe} are
compared to those obtained using the optimal analysis.}
\begin{center}
\begin{tabular}{|c|cc||c|cc|}
\hline
 \ & \multicolumn{2}{c||}{Technique} & \ & \multicolumn{2}{c|}{Technique} \\
 \ & $b_1$ \cite{gunhe} & optimal & \ & $b_1$ \cite{gunhe} & optimal \\
\hline
$D_1(B=0)/\sqrt S$ & 0.137 & 0.376 & $D_1(S=130,B=21)$ & 1.47 & 3.68 \\
$D_2(B=0)/\sqrt S$ & 0.663 & 1.68 & $D_2(S=130,B=21)$ & 7.10 & 16.5 \\
\hline
\end{tabular}
\end{center}
\label{discgamgam}
\end{table}

Consider, first, $F=\gam\gam$ in the limit of $B=0$. 
Table~\ref{discgamgam} gives $D_{1,2}(B=0)/\sqrt S$,
a first measure of the potential power for discrimination between
the models; we see a large improvement relative
to the $\ocp=b_1$ weighting procedure by using the optimal techniques.
Table~\ref{discgamgam} also shows the actual levels of discrimination
that can be achieved if we adopt
the LHC SM Higgs boson estimates of $S=130$ and $B=21$.
Of course, since the total number of events available is not large,
non-Gaussian effects in the statistics that are not incorporated
in our approach will prevent achieving results quite this good.

\begin{table}[h]
\caption[fake]{$t\anti tb\anti b$ channel, $\mh=100\gev$, $B/S=50$: 
signal event rates, and corresponding $S/\sqrt B$ values, 
required to achieve $D_{1,2}=2$
using the best $\ocp=b_4$ observable of Ref.~\cite{gunhe}
as compared to the optimal approach.}
\begin{center}
\begin{tabular}{|c|cc|cc|}
\hline
 \ & \multicolumn{2}{c|} {$D_1(B/S=50)=2$} & 
\multicolumn{2}{c|} {$D_2(B/S=50)=2$} \\
 \ & $S$ & $S/\sqrt B$ & $S$ & $S/\sqrt B$ \\
\hline
$b_4$ \cite{gunhe} & 15000 & 17 & 970 & 4.5 \\
optimal & 1592 & 5.64 & 80 & 1.26 \\
\hline
\end{tabular}
\end{center}
\label{discbb}
\end{table}

\begin{figure}[h]
\leavevmode
\begin{center}
\epsfxsize=4.25in
\hspace{0in}\epsffile{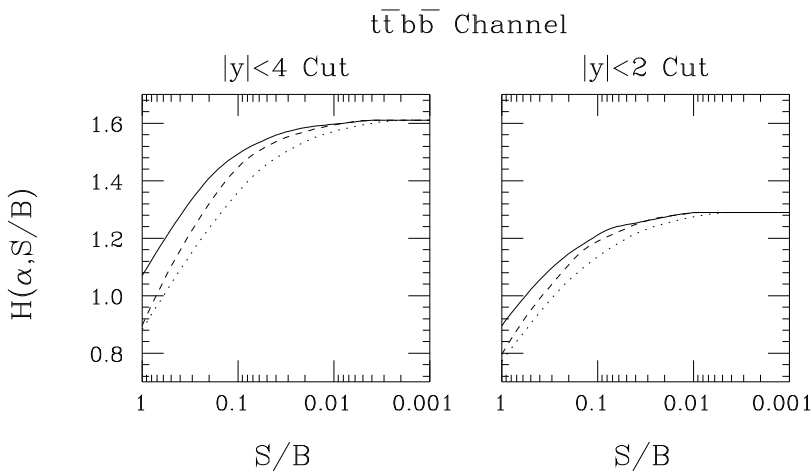}
\end{center}
\caption[]{We plot $H(\alpha,S/B)$ for the $t\anti t b\anti b$
final state as a function of $S/B$ for the three cases:
$\alpha=\alpha_I$ (solid); $\alpha=\alpha_{II}$ (dashes);
$\alpha=\alpha_{III}$ (dots).}
\label{hplots}
\end{figure}

We turn now to the $F=b\anti b$ channel.
The discrimination power $D$ for separating models is determined as a function
of $S/B$ using Eq.~(\ref{ddef})
and $H(\alpha_I,S/B)$ (computed for the $b\anti b$
channel) as plotted in Fig.~\ref{hplots}. 
In Table~\ref{discbb}, we give the number of signal events required to achieve
$D_{1,2}=2$ for $B/S=50$, and the corresponding $S/\sqrt B$ values. 
Also given are the results obtained using the best
$\ocp=b_4$ observable of Ref.~\cite{gunhe}. 
The improvement using the optimal techniques is very remarkable;
$D_{1,2}=2$ discrimination is possible with far fewer events
and much lower $S/\sqrt B$ as compared to Ref.~\cite{gunhe}.
For large $B/S$, the results in the $b\anti b$ final
state for $D_{1,2}$ are easily summarized:
\beq
D_1=C_1{S\over\sqrt B}\,,\quad D_2=C_2 {S\over\sqrt B}\,,
\label{d12summary}
\eeq
with $C_1=|\alpha_{II}-\alpha_{I}|H=0.36$
and   $C_2=|\alpha_{III}-\alpha_{I}|H=1.61$,
where $H=1.61$ is the result from Eq.~(\ref{hdef}).
These values of $C_1$ and $C_2$ 
can be compared to the values $0.091$ and $0.455$
achieved in Ref.~\cite{gunhe} using $\ocp=b_4$; the latter obviously
imply much weaker discrimination power.

A special remark is in order
regarding the small $S/\sqrt B=1.26$ value required
(see Table~\ref{discbb}) for $D_2=2$ in the $t\anti tb\anti b$
channel using the optimal approach.
If the Higgs boson mass is already known from observation in
another channel (\eg\ $t\anti t \gam\gam$), then one can 
perform the CP analysis for $t\anti t b\anti b$ events
falling into the appropriate $b\anti b$ mass bin (assuming mass calibrations
are well understood) even though a distinct bump is not observed.
In this way, one could  confirm and possibly improve
CP results obtained in the channel for which a mass bump is manifest.

Although the $\gam\gam$ channel is not likely to be useful for
Higgs bosons that do not have a large CP-even component, the $b\anti b$
channel is likely to be of very general utility. 
Light Higgs bosons will almost always have a large branching ratio
for decays to $b\anti b$. Given the large $t\anti t b\anti b$
background rate, $B/S$ will generally be large.
Thus, it is useful to give the $H(\alpha,S/B)$
values, from which the discrimination power $D$
of Eq.~(\ref{ddef}) can be computed, for other models.
The results for $H(\alpha_{II},S/B)$
and $H(\alpha_{III},S/B)$ appear along with those for $H(\alpha_I,S/B)$
in Fig.~\ref{hplots}. At $S/B=1/50$,
all are of order $\gsim 1.55$. They approach the common limit of $H=1.61$.
Using a lower bound of $H(\alpha,S/B)\gsim 1.5$ for $B/S>30$
(the most likely range),
we conclude that a marginal signal at the level of $S/\sqrt B=3$
will result in discrimination power of $D\geq 4.5|\alp-\alpha|$,
regardless of the value $\alpha$ for the input model.
This is quite an encouraging result in that for any Higgs boson that
is observable in the $t\anti t b\anti b$ channel one can hope
to determine the mix of CP-even and CP-odd Higgs couplings to $t\anti t$
at a very useful level.

\begin{table}[h]
\caption[fake]{Ratio $r=I_b/I_a$ of cross section contributions
arising from the CP-odd $b^2$ term and CP-even $a^2$ term
for various $|y|$ cut and $\mh$ (in GeV) choices.}
\begin{center}
\begin{tabular}{|c|c|c|c|c|}
\hline
Case & $|y|<4$ & $|y|<2$ & $|y|<4$ & $|y|<2$ \\
\ & $\mh=100$ & $\mh=100$ & $\mh=130$ & $\mh=130$ \\
\hline
 $r$ & 0.288 & 0.243 & 0.447 & 0.362 \\
\hline
\end{tabular}
\end{center}
\label{rsummary}
\end{table}

\begin{table}[h]
\caption[fake]{$t\anti t\gam\gam$ channel: $D_1$ and $D_2$ discrimination
powers obtained using the optimal technique, 
for the $|y|$ cut and $\mh$ values considered in Table~\ref{rsummary}.}
\begin{center}
\begin{tabular}{|c|c|c|c|c|}
\hline
Case & $|y|<4$ & $|y|<2$ & $|y|<4$ & $|y|<2$ \\
\ & $\mh=100$ & $\mh=100$ & $\mh=130$ & $\mh=130$ \\
\hline
$D_1(B=0)/\sqrt S$ & 0.376 & 0.292 & 0.449 & 0.336  \\
$D_1(S=130,B=21)$  & 3.68 & 3.07 & 4.31 & 3.50  \\
$D_2(B=0)/\sqrt S$ & 1.68 & 1.49 & 1.45 & 1.26  \\
$D_2(S=130,B=21)$  & 16.5 & 15.7 & 14.0 & 13.2  \\
\hline
\end{tabular}
\end{center}
\label{gamgamsummary}
\end{table}

We now explore sensitivity of these results to the $|y|$ cut
and to $\mh$. Since the $|y|<4$ rapidity cut for the $t$, $\anti t$ and 
$\h$ may be too generous to allow for full reconstruction
in the relatively complicated $t\anti t\h$ final state, we also consider
$|y|<2$.  We will explore sensitivity to $\mh$ by giving results for
$\mh=130\gev$.  First, we give the $r=I_b/I_a$ values, appropriate
in all the cases considered, in Table~\ref{rsummary}. Note that
as $\mh$ increases the overall signal cross section 
contribution ($I_b$) of the CP-odd $b^2$ term increases relative to the 
contribution ($I_a$) of the CP-even $a^2$ term, 
implying larger $r$ values at larger $\mh$.
Next, in Table~\ref{gamgamsummary} we give the discrimination powers
$D_{1,2}$ in the $\gam\gam$ mode. For fixed values of $S=130$ and $B=21$,
the stronger $|y|<2$ cut gives $D_{1,2}$ values that are not
so much smaller than for $|y|<4$.  However, this masks the fact that
only 1/3 as many events are accepted for $|y|<2$ as compared to $|y|<4$.  
Reducing $S$ and $B$ by a factor of 3 (for $|y|<2$) from $S=130$ and $B=21$
would yield $D_{1,2}$ values that are a factor of $1/\sqrt 3$
smaller than given in Table~\ref{gamgamsummary}.
Regarding $\mh$ dependence, note that,
keeping $S$ and $B$ constant, $D_1$ ($D_2$) is larger (smaller)
for $\mh=130\gev$ as compared to $\mh=100\gev$. The larger $D_1$
is because $|\alpha_{II}-\alpha_{I}|$ is larger (larger $r$) at
$\mh=130\gev$. This contrasts with the case of $D_2$ 
for which $|\alpha_{III}-\alpha_{I}|=1$ independent of $\mh$.

\begin{table}[h]
\caption[fake]{$t\anti tb\anti b$ channel: 
signal event rates, and corresponding $S/\sqrt B$ values, 
required to achieve $D_{1,2}=2$
using the optimal approach.}
\begin{center}
\begin{tabular}{|c|cc|cc|}
\hline
 \ & \multicolumn{2}{c|} {$D_1(B/S=50)=2$} & 
\multicolumn{2}{c|} {$D_2(B/S=50)=2$} \\
 ($|y|$ Cut,$\mh$) & $S$ & $S/\sqrt B$ & $S$ & $S/\sqrt B$ \\
\hline
(4,100) & 1592 & 5.64 & 80 & 1.26 \\
(2,100) & 3215 & 8.02 & 123 & 1.57 \\
(4,130) & 932 & 4.32   & 89 & 1.33 \\
(2,130) & 2018 & 6.35 & 143 & 1.69 \\
\hline
\end{tabular}
\end{center}
\label{bbsummary}
\end{table}

\begin{table}[h]
\caption[fake]{$t\anti tb\anti b$ channel: asymptotic $H$ values
and corresponding values of $C_1$ and $C_2$, as defined in
Eq.~(\ref{d12summary}).}
\begin{center}
\begin{tabular}{|c|c|c|c|c|}
\hline
Case & $|y|<4$ & $|y|<2$ & $|y|<4$ & $|y|<2$ \\
\ & $\mh=100$ & $\mh=100$ & $\mh=130$ & $\mh=130$ \\
\hline
$H$,$C_2$ & 1.61  & 1.29  & 1.53  & 1.20 \\
$C_1$     & 0.360 & 0.252 & 0.473 & 0.319 \\
\hline
\end{tabular}
\end{center}
\label{hcsummary}
\end{table}

Some comparative results in the $t\anti t b\anti b$ channel are
given in Tables~\ref{bbsummary} and \ref{hcsummary}. 
The $H(\alpha,S/B)$ values for $\mh=100\gev$ and $|y|<2$
were given in Fig.~\ref{hplots}. 
For $S/B=1/50$, one finds $H(\alpha_{I},1/50)=1.27$,
$H(\alpha_{II},1/50)=1.27$, and $H(\alpha_{III},1/50)=1.24$, a significant
deterioration from the $\gsim 1.55$ values found for $|y|<4$.  
The $S/B\to 0$ limit is $H=1.29$, again significantly smaller
than the $1.61$ value obtained with $|y|<4$ cuts.
More limited acceptance
implies that more signal rate (which is harder to achieve with
more limited acceptance, \eg\ $|y|<2$ accepts only 1/3 as many events
as $|y|<4$) and larger $S/\sqrt B$ is required for good discrimination
between the various models. Clearly it is
important to accept as much of the phase space as possible,
both to maximize the signal rate and to maximize the region
over which shape differences between models can be explored.
The trends with $\mh$ are similar to the $\gam\gam$ channel,
with $D_1$ ($D_2$) being larger (smaller) for $\mh=130\gev$ as compared
to $\mh=100\gev$, for the same $S/\sqrt B$.

\section{The MSSM Higgs Bosons}

A Higgs boson sector of particular interest is that of the minimal
supersymmetric model (MSSM) \cite{hhg}. 
If the mass of the CP-odd $\ha$ of the model
is $> 150\gev$, then the light CP-even $\hl$ will be SM-like, and
our $t\anti t\h$ analysis will apply in both the $\gam\gam$ and $b\anti b$
channels. However, the heavier CP-even $\hh$ and the $\ha$ 
will probably present a greater 
challenge.  If the $\tanb$ parameter of the MSSM is near 1, then
the $t\anti t\hh,t\anti t\ha$ couplings, and corresponding LHC 
production rates, will be substantial for modest $\mhh,\mha$.
Depending upon how easily they can be discovered in the relevant 
decay channels (such as $\hh,\ha\to b\anti b,\tauptaum$, $\hh\to\hl\hl$,
$\ha\to\hl Z$, and, for $\mhh,\mha\geq 2\mt$, 
$\hh,\ha\to t\anti t$)~\footnote{Their $\gam\gam$ 
branching ratios are typically too small to be useful.} 
and upon how small $S/B$ in each channel is, our $t\anti t\h$
procedures could be useful.  We do not attempt a detailed study here. 

If the $\tanb$ parameter of the model is large
and $\mha\sim\mhh> 150\gev$, then
the $t\anti t\hh$ and $t\anti t \ha$ couplings 
are suppressed by a factor of $1/\tanb$ and those to $b\anti b$
(and $\tauptaum$) are enhanced by a factor of $\tanb$.  As a result,
the primary LHC production modes are 
$pp\to b\anti b\hh X,b\anti b\ha X$,
where both $\hh$ and $\ha$ decay to $b\anti b$ ($\tauptaum$) with a branching
fraction of order 0.9 (0.1). In this case, only the 
$b\anti b b\anti b$ and $b\anti b \tauptaum$
final states are relevant for detecting the $\hh$ and $\ha$.
Both ATLAS \cite{ATLAS} and CMS \cite{CMS}
claim that $S/\sqrt B=5$ discovery of the $\hh,\ha$
is possible (for full `three-year' $L=300\fbi$ integrated luminosity
per detector)
in the $b\anti b\tauptaum$ mode provided $\tanb$ is above a
$\mha$-dependent lower limit ranging from $\tanb\sim 2$
at $\mha\sim\mhh\sim 150\gev$ up to $\tanb\sim 20$ at $\mha\sim\mhh\sim
500\gev$. Prospects in the $b\anti b b\anti b$ final state 
are more controversial. The $b\anti b\tauptaum$ mode is preferable for
our purposes in any case because the $\tau$'s allow us to identify
the Higgs boson without any combinatoric uncertainty. The question,
then, is whether or not our techniques will allow verification
of the CP$=+$ and CP$=-$ nature of the $\hh$ and $\ha$, respectively. 

First, and very crucially, we do not need to identify
$b$ vs. $\anti b$ due to the fact (see earlier) that
$f_a(\phi)$ and $f_b(\phi)$ are independent
under interchange of the $b$ and $\anti b$ momenta.
However, there are two significant difficulties.
First, in Eq.~(\ref{sigform}) $f_a(\phi)\simeq f_b(\phi)$
in the limit where the quark mass approaches
zero. (That is, $\Sigma(\phi)\propto a^2+b^2$ in this limit,
the $a^2-b^2$ term being suppressed by $\sim m_{\rm quark}/\sqrt{\hat s}$
in comparison.) Consequently, a very large number of events will be required
for a statistically useful determination of the $a^2-b^2$ coefficient.
Second, mass resolution in the $\tauptaum$ (or $b\anti b$)
channel will typically not be adequate for separating the $\hh$ and $\ha$
mass peaks, implying that one could at best verify that $a^2-b^2=0$
after averaging over the $\hh$ and $\ha$ (given
that the $\hh$ and $\ha$ have the same production rate at large $\tanb$). 
Our first estimates suggest that,
for $\tanb$ values such that discovery is possible,
a statistically meaningful verification
of the $a^2=b^2$ expectation relative to pure $a^2=0$
or pure $b^2=0$ might be possible, 
despite the small $m_b/\sqrt{\hat s}$ coefficient multiplying $a^2-b^2$.
A careful study of this issue by the detector collaborations --- \ie\ one that
includes all resolutions, minimum-bias backgrounds, and so forth ---
is needed to properly assess the prospects.

\section{Final Remarks and Conclusion}

In this paper, we have employed the optimal analysis procedures
of Ref.~\cite{optimal} to determine the accuracy with which
the relative magnitude of the CP-even and CP-odd $t\anti t$
couplings of a light Higgs boson can be determined 
using the final state kinematical distributions of the $t$, $\anti t$
and $\h$ relative to one another in LHC $pp\to t\anti t \h$ events.
We have found that the optimal analysis procedure
yields a large improvement in the statistical discrimination power
($D\equiv \sqrt{\chi^2}$) 
as compared to the simpler weighting function techniques previously
considered in Ref.~\cite{gunhe}. 

We emphasize that we have employed 
a discrimination measure that does not make use
of the (substantial) sensitivity of the absolute production rate 
to the CP-even versus CP-odd couplings; this is desirable since
absolute rates have additional systematic uncertainties
associated with model dependence for branching ratios, experimental
efficiencies and the like.
However, an estimate or, preferably, experimental
measurement of the signal to background ratio is needed
in order to compute quantitative results for
the $\chi^2$ associated with a model that is different
from the input or true model.

It is useful to note that in the actual experimental analysis
an equivalent of the optimal analysis procedure employed here would be to
determine the likelihood of the fit, as a function of the $a$
and $b$ parameters, to the detailed phase space distribution (allowing the
overall normalization to be a free parameter). In the limit of Gaussian
statistics, this and the theoretically much more convenient optimal
procedure employed here are equivalent: the ratio of the likelihood
of a fit in an incorrect model to that in the correct model
will be equal to $\exp[-\chi^2]$, where the $\chi^2$
is the $\chi^2$ of an incorrect model choice 
as computed in the optimal procedure using inputs based on the correct model.

Our conclusion is that 
if a Higgs signal is observable in the $t\anti t\h$ ($\h\to F$, $F=b\anti b$
or $\gam\gam$) final states at the LHC, then the relative size
of the CP-even and CP-odd couplings of the $\h$ to $t\anti t$
can be determined at a statistically useful level.
In particular, in the case of a SM-like Higgs boson,
we find that Higgs CP coupling mixtures that are significantly different
from pure CP-even can generally be excluded at a high level of
statistical significance.

Finally, we noted that there is reason to hope
that the techniques studied here
could be employed for studying the CP nature of the heavier Higgs
bosons of the minimal supersymmetric model. In particular, for $\tanb$
values large enough that a signal is observable in the $b\anti b \tauptaum$
final state coming from $b\anti b\hh$ and $b\anti b\ha$ production 
(with $\hh,\ha\to\tauptaum$), one can hope to verify that the overlapping
$\hh$ and $\ha$ resonances yield the expected 
equal mixture of CP-even and CP-odd signals.

\bigskip
\centerline{\bf Acknowledgements}
\medskip
This work was supported in part by Department of Energy under
grant No. DE-FG03-91ER40674,
by the Davis Institute for High Energy Physics, by the
State Committee for Scientific Research (Poland)
under grant No. 2 P03B 014 14, and by Maria Sk\l odowska-Curie
Joint Fund II (Poland-USA) under grant No. MEN/NSF-96-252.
JP would like to thank the Davis Institute for High Energy Physics for
hospitality. We would like to thank B. Grzadkowski for helpful conversations.

\end{document}